\RequirePackage{fix-cm}
\documentclass[a4paper,11pt]{article}

\usepackage{authblk}
\usepackage{amsfonts,amsmath,amssymb,amsthm}
\usepackage{algorithm}
\usepackage{paralist}
\usepackage{booktabs}
\usepackage{url}
\usepackage{hyperref}
\usepackage{graphicx}
\usepackage{microtype}
\usepackage{url}
\usepackage{hyperref}
\usepackage{subfig}
\usepackage{tikz}
\usetikzlibrary{positioning}
\usetikzlibrary{decorations.pathreplacing}

\newcommand{\N}{\mathbb{N}}

\newcommand{\F}{\mathbb{F}}

\newtheorem{definition}{Definition}
\newtheorem{lemma}{Lemma}
\newtheorem{theorem}{Theorem}

\tikzset{
    mybrace/.style={decorate,decoration={brace,aspect=#1}}
}

\providecommand{\keywords}[1]{\textbf{\textit{Keywords }} #1}

\begin{document}

\title{Heuristic Search of (Semi-)Bent Functions based on Cellular Automata}

\author[1]{Luca Mariot}
\author[2]{Martina Saletta}
\author[2]{Alberto Leporati}
\author[3]{Luca Manzoni}
	
\affil[1]{{\small Cyber Security Research Group, Delft University of Technology, Mekelweg 2, Delft, The Netherlands} 
	
	{\small \texttt{l.mariot@tudelft.nl}}}

\affil[2]{{\small Dipartimento di Informatica, Sistemistica e Comunicazione, Università degli Studi di Milano-Bicocca, Viale Sarca 336/14, Milano, 20126, Italy}
    	{\small \texttt{\{martina.saletta,alberto.leporati\}@unimib.it}}}

\affil[3]{{\small Dipartimento di Matematica e Geoscienze, Università degli Studi di Trieste, Via Valerio 12/1, Trieste, 34127, Italy}

    {\small \texttt{lmanzoni@units.it}}}

\maketitle

\begin{abstract}
An interesting thread in the research of Boolean functions for cryptography and coding theory is the study of \emph{secondary constructions}: given a known function with a good cryptographic profile, the aim is to extend it to a (usually larger) function possessing analogous properties. In this work, we continue the investigation of a secondary construction based on cellular automata, focusing on the classes of bent and semi-bent functions. We prove that our construction preserves the algebraic degree of the local rule, and we narrow our attention to the subclass of quadratic functions, performing several experiments based on exhaustive combinatorial search and heuristic optimization through Evolutionary Strategies (ES). Finally, we classify the obtained results up to permutation equivalence, remarking that the number of equivalence classes that our CA-XOR construction can successfully extend grows very quickly with respect to the CA diameter.
\end{abstract}


\section{Introduction}
\label{sec:intro}
The design of symmetric ciphers traditionally revolves around the concepts of confusion and diffusion, introduced by Shannon~\cite{shannon49} as a general guideline to develop encryption functions that are able to frustrate statistical attacks. Specifically, \emph{confusion} prescribes that the relationship between the ciphertext and the encryption key should be as complicated as possible, while the aim of \emph{diffusion} is to spread the statistical structure of the plaintext over the ciphertext---or equivalently, to make the value of every ciphertext bit depend on many plaintext bits (ideally all of them).

The study of diffusion layers in block ciphers mainly leverage on the theory of error-correcting codes to come up with transformations that propagate the differences of very similar plaintexts in an optimal way (see e.g. the {\sc Mix-Columns} operation in AES~\cite{daemen20}, which relies on an MDS matrix). On the other hand, \emph{Boolean functions} (i.e. mapping of the type $f: \{0,1\}^n \to \{0,1\}$) with specific properties are usually sought to design confusion layers, for instance in the form of \emph{combining} or \emph{filtering} functions in stream ciphers~\cite{carlet21}. In block ciphers, \emph{S-boxes} (i.e. the vectorial generalization of Boolean functions) are used to design the confusion layer instead, as in the \emph{Substitution-Permutation Network} paradigm~\cite{stinson18}. The rationale behind choosing a particular Boolean function when designing a symmetric cipher usually boils down to check several \emph{cryptographic properties}, which are aimed at countering specific attacks.

For the above reason, research in the past decades focused on methods for the construction of Boolean functions with good cryptographic properties, about which the reader can find an excellent account in Carlet's recent book~\cite{carlet21}. These methods are usually classified in two approaches. In \emph{primary constructions}, one seeks to construct Boolean functions with certain properties starting ``from scratch'', i.e., by leveraging on other combinatorial objects such as permutations (see the \emph{Maiorana-McFarland construction}~\cite{mcfarland73}) or finite vector spaces (as in Dillon's \emph{partial spread construction}~\cite{dillon74}). On the other hand, a \emph{secondary construction} starts from an existing Boolean function with a certain cryptographic profile, and constructs a new one (usually over a larger number of variables) possessing similar properties. The best known example in this respect is \emph{Rothaus's construction}~\cite{rothaus76}.

\emph{Cellular Automata} (CA) have been extensively investigated as a computational building block for designing several cryptographic primitives. The most famous examples include Wolfram's pseudorandom number generator for Vernam-like stream ciphers~\cite{wolfram85}, which was based on the chaotic dynamics of the elementary local rule 30 (later shown to be vulnerable to correlation and approximation attacks~\cite{meier91,koc97}), and the $\chi$ nonlinear transformation by Daemen et al.~\cite{daemen94}, used as an S-box in {\sc Keccak}~\cite{bertoni11} and other symmetric ciphers. As far as we know, most of these works focused on the study of certain local rules that, when plugged into a CA, would yield good cryptographic functions. Such studies, however, usually rely on ad-hoc arguments that mostly depend on the specificities of the single rules, and are hardly generalizable. To the best of our knowledge, there have been no significant attempts to use CA as a general method to define primary or secondary constructions of Boolean functions with good cryptographic properties. The only exception we are aware of is the use of linear bipermutive CA as a primary construction for \emph{bent} Boolean functions belonging to the partial spreads class~\cite{gadouleau20}, recently proposed by the first author of this paper together with M. Gadouleau and S. Picek.

In this paper, we continue the investigation of a secondary construction of Boolean functions set forth in the paper ``\emph{Exploring Semi-bent Boolean Functions Arising from Cellular Automata}''~\cite{mariot20a}, presented at ACRI 2020. The idea underlying the construction, which we name the \emph{CA-XOR construction}, is quite simple: a CA of $n$ cells is evolved for a single time step, and the XOR of the output cells is taken as the value of a Boolean function of $n$ variables, where the CA initial configuration constitutes the input vector. The properties of the resulting function depend on the underlying local rule used to evolve the CA, which is itself a Boolean function of $d \le n$ variables. The goal is thus to start from a Boolean function with good cryptographic properties such that, when used as a local rule in the CA-XOR construction, results in a larger function with similar good properties. Hence, we actually obtain a \emph{recursive} secondary construction with our method: indeed, one can define an infinite family of functions by just adding enough cells to the CA. In particular, given a starting function of $d$ variables, it is possible to define a new function for each $n > d$.

The present manuscript extends the work in~\cite{mariot20a} along several directions:
\begin{itemize}
\item We enlarge the scope of our investigation by considering also bent functions, instead of only semi-bent ones as in the original paper. Since bent functions exist only when the number of variables $n$ is even, we focus our attention on those (semi-)bent functions that, when used as a starting point in our CA-XOR construction, always yield bent functions when the number of CA cells $n$ is even, and semi-bent functions when $n$ is odd.
\item We refine the computational search approach laid down in~\cite{mariot20a}. By leveraging on some known results on quadratic bent functions, we simplify the combinatorial algorithm by enumerating only the set of \emph{homogeneous} algebraic normal forms of degree $2$, i.e. those without linear terms. This allows us to extend our exhaustive search experiments up to diameter $d=7$.
\item For higher diameters where exhaustive search becomes unfeasible, we devise an \emph{Evolutionary Strategies} (ES) optimization algorithm. Exploiting the fact that there exists only one (semi-)bent Boolean function of $d$ variables up to \emph{Extended Affine} (EA)-\emph{equivalence}, we use ES to evolve affine transformations applied to this function. The optimization objective thus becomes to find a suitable affine transformation such that the resulting quadratic function can be extended by the CA-XOR construction up to $n=16$ cells.
\item Finally, we gather all the functions obtained through the exhaustive search and ES optimization experiments, and classify them up to \emph{permutation equivalence}~\cite{carlet21}. We report the numbers of the obtained equivalence classes, observing that they grow very quickly with respect to the diameter of the CA.
\end{itemize}

The remainder of this paper is structured as follows. Section~\ref{sec:prelim} recalls the basic definitions and results concerning Boolean functions and cellular automata used throughout the paper. Section~\ref{sec:rel-works} gives a general overview of the literature concerning the use of CA in symmetric cryptography, and covers the main works related to the algebraic and heuristic constructions of Boolean functions with good cryptographic properties. Section~\ref{sec:constr} introduces the CA model considered in this work and defines the CA-XOR construction of Boolean functions, proving some basic theoretical facts about it. Section~\ref{sec:exhaust} describes the search algorithm used to exhaustively enumerate quadratic functions with a homogeneous algebraic normal form, while Section~\ref{sec:es} is devoted to the Evolutionary Strategies optimization algorithms used to evolve quadratic functions on larger diameters. Section~\ref{sec:class} presents the results of our experiments, providing a classification of the obtained functions up to permutation equivalence. Section~\ref{sec:outro} summarizes the key contributions of the paper, and discusses several open problems for future research on the subject.

\section{Preliminary Definitions}
\label{sec:prelim}
In this section, we gather all background definitions related to Boolean functions and cellular automata used in the following. As a general notation, let $\F_2 = \{0,1\}$ be the finite field of two elements, where the sum operation is the XOR (denoted by $\oplus$) while multiplication is the logical AND (denoted by simple concatenation). The $n$-dimensional vector space over $\F_2$ (i.e., the set of all $n$-bit vectors) is denoted $\F_2^n$. The \emph{support} of $x \in \F_2^n$ is defined as $supp(x) = \{i : x_i \neq 0 \}$, while the \emph{Hamming weight} of $x$ is $w_H(x)=\vert supp(x) \vert$, i.e. the number of $1$s in $x$. The vector space $\F_2^n$ is usually equipped with the \emph{scalar product} defined as $a\cdot x = \bigoplus_{i=1}^n a_ix_i$ for all $a,x \in \F_2^n$. Further, for all $n \in \N$ we denote by $[n] = \{1,\cdots, n\}$ the set of the first $n$ positive integer numbers.

\subsection{Boolean Functions}
\label{subsec:bf}
The body of literature devoted to Boolean functions for cryptography and coding theory is quite extensive. In this section, we only recall the essential notions related to the main cryptographic properties of Boolean functions that we will use in the rest of the paper, referring the reader to~\cite{carlet21} for a more thorough treatment of the topic.

A \emph{Boolean function} of $n \in \N$ variables is a mapping $f: \F_2^n \to \F_2$, i.e. a function from the set of $n$-bit vectors to a single bit. There exist several ways to represent Boolean functions, the simplest one being the \emph{truth table}. Assuming that $\F_2^n$ is endowed with a certain total order (e.g., the lexicographic order), the truth table of a function $f: \F_2^n \to \F_2$ is the $2^n$-bit vector $\Omega_f \in \F_2^{2^n}$ that specifies the output value of $f$ for each of the vectors in $\F_2^n$. The \emph{weight} of $f$ is defined as the Hamming weight of its truth table $\Omega_f$. Further, function $f$ is called \emph{balanced} if $w_H(\Omega_f) = 2^{n-2}$, that is, if its truth table is composed of an equal number of 0s and 1s. Balancedness is a fundamental cryptographic property for Boolean functions used in stream and block ciphers.

Although being the simplest representation, most of the cryptographic properties of a Boolean function are difficult to define in terms of its truth table. To this end, another more useful unique representation is the \emph{Algebraic Normal Form} (ANF). Remarking that $x^2 = x$ for all $x \in \F_2$, the ANF of a Boolean function $f: \F_2^n \to \F_2$ is defined as the following multivariate polynomial over the quotient ring $\mathbb{F}_2[x_1,\cdots,x_n]/(x_1^2 \oplus x_1, \cdots, x_n^2 \oplus x_n)$:
\begin{equation}
    P_f(x) = \bigoplus_{I \in 2^{[n]}} a_I \left( \prod_{i \in I} x_i \right) \enspace ,
\end{equation}
where $2^{[n]}$ is the power set of $[n] = \{1,\cdots,n\}$. The \emph{algebraic degree} of $f$ is the cardinality of the largest subset $I \in 2^{[n]}$ in its ANF such that $a_I \ne 0$, or equivalently the size of its largest nonzero monomial. As a cryptographic criterion, the algebraic degree should be as high as possible. \emph{Affine functions} are defined as those Boolean functions with degree at most $1$. In particular, the ANF of an affine function is defined as a scalar product between a fixed vector $a \in \F_2^n$ and the input vector $x \in \F_2^n$, plus a constant $b \in \F_2$, i.e. $A_{a,b}(x) = a\cdot x \oplus b$. If $b=0$, then the function is called $linear$. Thus, there exist $2^n$ linear functions of $n$ variables. In a similar way, \emph{quadratic functions} are the Boolean functions of degree at most $2$. The fact that ``at most'' is used instead of ``exactly'' is for simplicity of language when talking about the derivatives of Boolean functions (see~\cite{carlet21}, page 53 for further explanations). Further, the set of all Boolean functions of degree at most $d$ has a vector space structure, since it is closed under sum.

 The vector of the ANF coefficients $a_I$ and the truth table of $f$ are related through the \emph{M{\"o}bius transform}:
\begin{equation}
  \label{eq:mobius}
  f(x) = \bigoplus_{I \in 2^{[n]}: I \subseteq supp(x) } a_I \enspace .
\end{equation}
In particular, the \emph{M{\"o}bius transform} is an \emph{involution}, meaning that one can apply the same formula in Equation~\eqref{eq:mobius} to recover the vector of ANF coefficients from the truth table of a function.

A third way to uniquely represent Boolean functions is the \emph{Walsh transform}, which captures several cryptographic properties. Formally, the Walsh transform of a Boolean function $f: \F_2^n \to \F_2$ is defined for all $a \in \F_2^n$ as:
\begin{equation}
    \label{eq:walsh}
    W_f(a) = \sum_{x \in \F_2^n} (-1)^{f(x) \oplus a\cdot x} \enspace .
\end{equation}
A function $f$ is balanced if and only if the Walsh coefficient over the null vector is zero, i.e. $W_f(0) = 0$. More in general, the coefficient $W_f(a)$ measures the \emph{correlation} between $f$ and the linear function $a\cdot x$. Thus, the Walsh transform can be used to compute the \emph{nonlinearity} of a Boolean function $f$, which is defined as the minimum Hamming distance of (the truth table of) $f$ from the set of all affine functions. More precisely, the nonlinearity of $f$ equals
\begin{equation}
    \label{eq:nl}
N_f = 2^{n-1} - \frac{1}{2}\cdot \max_{a \in \F_2^n}\{ \vert W_f(a) \vert \} \enspace .
\end{equation}
The nonlinearity of Boolean functions used in the combiner and filter model of stream ciphers should be as high as possible~\cite{carlet21}. From Equation~\eqref{eq:nl}, this means that the maximum absolute value of the Walsh transform should be as low as possible. By \emph{Parseval relation}, this can happen only when all Walsh coefficients have the same absolute value $2^{\frac{n}{2}}$, yielding the \emph{covering radius bound}: $N_f \le 2^{n-1} - 2^{\frac{n}{2} - 1}$. Functions satisfying this bound are called \emph{bent}, and they exist only when $n$ is even (since the Walsh transform can yield only integer numbers). Such functions are not balanced, since $W_f(0) = \pm 2^{\frac{n}{2}}$, and thus they cannot be used directly in the design of stream and block ciphers. However, there exist several ways to construct highly nonlinear balanced functions from bent ones (such as, for example, XORing the value of the function with an independent additional variable). For $n$ odd, the \emph{quadratic bound} is given by $N_f \le 2^{n-1} - 2^{\frac{n+1}{2} - 1}$. The name of this bound comes from the fact that it can be always achieved by functions of algebraic degree $2$. In general, finding the highest possible nonlinearity between the quadratic and the covering radius bound is an open problem for $n>7$ odd.

\emph{Plateaued functions} represent an interesting generalization of bent functions, since they can also be balanced while still retaining high nonlinearity. Formally, a Boolean function $f: \F_2^n \to \F_2$ is \emph{plateaued} if its Walsh transform takes only three values, i.e. if $W_f(a) \in \{-\lambda, 0, +\lambda\}$ for all $a \in \F_2^n$. In particular, a plateaued function is \emph{semi-bent} if $\lambda = 2^{\frac{n+1}{2}}$ for $n$ odd and $\lambda = 2^{\frac{n+2}{2}}$ for $n$ even.  This means that the nonlinearity of a semi-bent function equals $2^{n-1} - 2^{\frac{n-1}{2}}$ when $n$ is odd and $2^{n-1} - 2^{\frac{n}{2}}$ when $n$ is even. Hence, semi-bent functions reach the quadratic bound for nonlinearity when $n$ is odd.

We conclude this section by giving a short overview of the \emph{equivalence relations} studied in the context of Boolean functions. The size of the space of $n$-variable Boolean functions is $2^{2^n}$, i.e. it is superexponential in $n$. Therefore, one cannot perform an exhaustive search of all Boolean functions to find those with the desired cryptographic properties for a specific application already for $n>5$. Using equivalence relations that preserve these properties helps in reducing the search space size, by looking only at the set of equivalence classes. The coarsest equivalence relation used for Boolean functions is \emph{permutation equivalence}. Namely, two functions $f,g: \F_2^n \to \F_2$ are called permutation equivalent if there exists a permutation $\pi: [n] \to [n]$ such that
\begin{equation}    
\label{eq:perm-eq}
f(x_1, x_2, \cdots, x_n) = g(x_{\pi(1)}, x_{\pi(2)}, \cdots, x_{\pi(n)})
\end{equation}
for all $x \in \F_2^n$. In simpler terms, $f$ and $g$ are permutation equivalent if it is possible to reorder the input variables of $g$ to match the output of $f$. A more general relation, which we will use extensively in this paper, is affine equivalence. In this case, we say that $f$ and $g$ are affinely equivalent if there exist an invertible $n\times n$ binary matrix $M$ and a vector $v \in \F_2^n$ such that
\begin{equation}
\label{eq:aff-eq}
f(x) = g(Mx^\top \oplus v) 
\end{equation}
for all $x \in \F_2^n$. In this case, the two functions are equivalent if by applying the affine transformation $Mx^\top \oplus v$ to the input vector $x$ and then evaluating $g$ on the result, one gets the output of $f$ computed directly on $x$. When $v=\underbar{0}$ is the null vector we say that the two functions are \emph{linearly equivalent}. Remark that permutation equivalence is a special case of linear equivalence, since one can define a permutation of the input variables in terms of a \emph{permutation matrix}. Concerning the cryptographic properties, affine equivalence (and therefore, linear and permutation equivalence which are special cases) preserves the algebraic degree and the nonlinearity.

Two further equivalence relations are usually studied in the context of Boolean functions, namely \emph{Extended Affine} (EA) equivalence and \emph{Carlet-Charpin-Zinoviev} (CCZ) equivalence. However, we will not cover them here, since CCZ equivalence coincides with EA equivalence for single-output Boolean functions (i.e. it is relevant only for vectorial Boolean functions, or S-boxes), while EA-equivalence does not preserve the algebraic degree. In particular, we are interested in the theorem below, whose proof can be found in~\cite{carlet21}:

\begin{theorem}
\label{thm:equiv-quad}
The following results hold:
\begin{itemize}
\item If $n$ is even, every bent quadratic function of $n$ variables is affinely equivalent to the function
\begin{equation}
\label{eq:bent-quad}
b(x) = x_1x_2 \oplus x_2x_3 \oplus \cdots x_{n-1}x_n \enspace .
\end{equation}
\item If $n$ is odd, every semi-bent quadratic function of $n$ variables is affinely equivalent to the function
\begin{equation}
\label{eq:sbent-quad}
s(x) = x_1x_2 \oplus x_2x_3 \oplus \cdots x_{n-2}x_{n-1} \oplus x_{n} \enspace .
\end{equation}
\end{itemize}
\end{theorem}
Therefore, there exist only one (semi-)bent function of $n$ variables up to affine equivalence, whose ANF is the sum of all separate binomials in increasing order (excluding the last variable which is summed independently for semi-bent functions on an odd number of variables).

\subsection{Cellular Automata}
\label{subsec:ca}
Cellular Automata (CA) are one of the oldest nature-inspired computational models studied in the literature. Indeed, they were initially considered by Ulam~\cite{ulam52} and Von Neumann~\cite{vonneumann66} in the 1950-60s to investigate self-replication phenomena. Nowadays, CA are applied both as a simulation tool for complex systems and as a computing device in the most disparate domains, ranging from physics to ecology. In essence, a CA is characterized by a regular lattice of \emph{cells} which can be in a discrete set of values. The global state of the system is updated by the synchronous application of a \emph{shift-invariant transformation}. In other words, each cell applies the same \emph{local rule} in parallel to decide its next state. Such rule is evaluated on the configuration of states formed by the cell itself and its neighboring cells. The size of the neighborhood is determined by the \emph{diameter} of the CA.

Usually, research on cellular automata focuses on the study of their \emph{long-term} dynamic behavior, which emerges from repeatedly applying the local rule on each cell for multiple time steps. On the contrary, in this work we adopt a different viewpoint: since we are mainly interested in CA as a layer to implement meaningful cryptographic primitives, we focus only on their \emph{short-term} behavior. This implies, in particular, analyzing the algebraic properties of the functions induced by the application of the CA local rule over a \emph{single} time step. Although there are approaches in the literature that consider the synthesis of Boolean circuits by evolving the CA for multiple steps (see e.g.~\cite{hazari18}), we decided to employ this setting for two reasons: first, it greatly simplifies the theoretical analysis of the cryptographic primitives defined by CA. Second, several examples of symmetric ciphers that exploits CA adopt this design direction. In particular, past experience seems to indicate that the best way to employ CA in a symmetric cipher is as nonlinear transformations applied for a single time step, and combining their output with that of other non-CA components to improve diffusion. The best known example of this design pattern is the $\chi$ transformation used in {\sc Keccak}~\cite{bertoni11}, which is now part of the SHA-3 standard for cryptographic hash functions.

We now introduce the basic CA model considered in the rest of this work.
\begin{definition}
\label{def:ca}
A \emph{No-Boundary Cellular Automaton} (NBCA) of length $n$ and diameter $d\le n$ is a vectorial Boolean function $F: \F_2^n \to \F_2^{n-d+1}$ defined for all $x = (x_1, x_2, \cdots, x_n) \in \F_2^n$ as:
\begin{equation}
\label{eq:ca}
F(x_1, x_2, \cdots, x_n) = (f(x_1, \cdots, x_d), \cdots, f(x_{n-d+1}, \cdots, x_n)) \enspace ,
\end{equation}
where $f: \F_2^d \to \F_2$ is a Boolean function of $d$ variables called the CA \emph{local rule}.
\end{definition}
\noindent
Hence, a NBCA can be seen as a lattice of $n$ cells arranged over a line, where each cell $i \in [n-d+1]$ computes its output state by applying the local rule $f$ on itself and the $d-1$ right neighbors $\{i+1,\cdots i+d-1\}$. The rightmost $d-1$ cells are effectively lost after the application of the global rule $F$, since they do not have enough right neighbors to compute their next state. Although there are methods in the literature to address this issue (e.g. with \emph{null} or \emph{periodic} boundary conditions), in this work we are not interested in keeping the number of cells constant, since as we argued above we only consider a single application of the CA global rule $F$. Remark that the NBCA model as stated in Definition~\ref{def:ca} has also been studied for other cryptographic applications, namely in~\cite{mariot19} for CA-based S-boxes, and in~\cite{mariot20} for mutually orthogonal Latin squares.

\begin{figure}[t]
    \centering
    \subfloat[CA 90 evaluated over 00110100.\label{fig:nbca-a}]{
        \centering
    \begin{tikzpicture}
    [->,auto,node distance=1.5cm, empt node/.style={font=\sffamily,inner
        sep=0pt}, rect
    node/.style={rectangle,draw,thick,font=\bfseries\sffamily,minimum size=0.7cm, inner
        sep=0pt, outer sep=0pt},
        grey node/.style={rectangle,draw,thick,fill=gray!30,font=\bfseries\sffamily,minimum size=0.7cm, inner
        sep=0pt, outer sep=0pt}]
    
    \node [empt node] (c)   {};
    \node [grey node] (c1) [right=0.1cm of c] {1};
    \node [grey node] (c2) [right=0cm of c1] {1};
    \node [grey node] (c3) [right=0cm of c2] {1};
    \node [rect node] (c4) [right=0cm of c3] {0};
    \node [rect node] (c5) [right=0cm of c4] {0};
    \node [grey node] (c6) [right=0cm of c5] {1};
    
    \node [empt node] (f1) [above=0.5cm of c2.east] {{\footnotesize
            $f(0,0,1) = 1$}};
    
    \node [rect node] (p2) [above=0.85cm of c1] {0};
    \node [rect node] (p1) [left=0cm of p2] {0};
    \node [grey node] (p3) [right=0cm of p2] {1};
    \node [grey node] (p4) [right=0cm of p3] {1};
    \node [rect node] (p5) [right=0cm of p4] {0};
    \node [grey node] (p6) [right=0cm of p5] {1};
    \node [rect node] (p7) [right=0cm of p6] {0};
    \node [rect node] (p8) [right=0cm of p7] {0};
    
    \node [empt node] (p7) [below=0.2cm of p1] {};
        \node [empt node] (p7a) [below=0.5cm of c1] {\phantom{M}};
    \node [empt node] (p8) [right=0.07cm of p7] {};
    \node [empt node] (p12) [above=0.5cm of p1.east] {};
    \node [empt node] (p13) [above=0.5cm of p5.east] {};
    \node [empt node] (p14) [above=0.3cm of p13] {\phantom{M}};
    
    \draw [-,thick, mybrace=0.25, decorate, decoration={brace,mirror,amplitude=5pt,raise=0.4cm}]
    (p1.west) -- (p3.east) node [midway,yshift=-0.3cm] {};
    (p1.west) -- (p2.east) node [midway,yshift=0.3cm] {};
    \draw[->,thick,shorten >=0pt,shorten <=0pt,>=stealth] (p8) -- (c1.north);
    \draw[->, draw=white] (p12) edge[bend left] (p13);
      \end{tikzpicture}
    }%
    \subfloat[Truth table of rule 90.\label{fig:nbca-b}]{
      \begin{tabular}{cc}
            \hline\noalign{\smallskip}
            $x_i,x_{i+1},x_{i+2}$ & $f(x_i,x_{i+1},x_{i+2})$ \\
            \noalign{\smallskip}\hline\noalign{\smallskip}
            000 & 0 \\
            100 & 1 \\
            010 & 0 \\
            110 & 1 \\
            001 & 1 \\
            101 & 0 \\
            011 & 1 \\
            111 & 0 \\
            \hline\noalign{\smallskip}
          \end{tabular}
    }
    \caption{Example of CA of $n=8$ cells equipped with rule 90 of diameter $d=3$, defined as $f(x_i,x_{i+1},x_{i+2})=x_i \oplus x_{i+2}$.}
    \label{fig:nbca}
\end{figure}
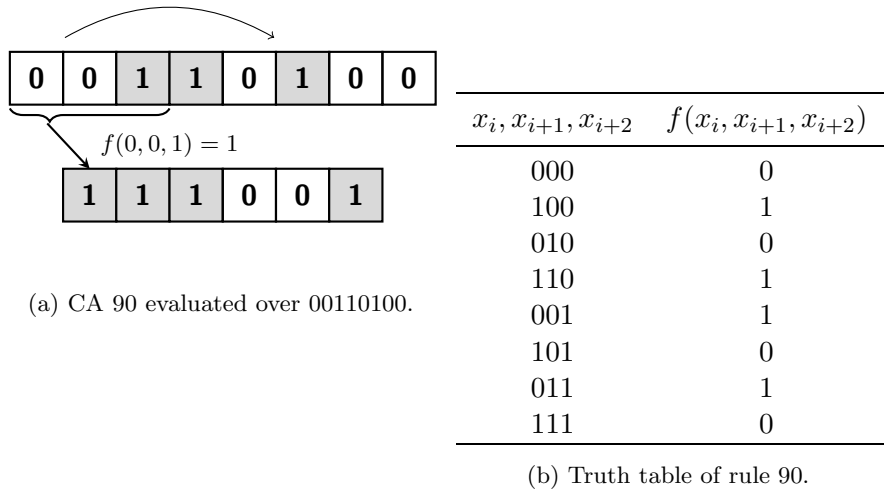

Since the local rule $f: \F_2^{d} \rightarrow \F_2$ is a Boolean function, it can be defined by a truth table $\Omega_f$ of $2^{d}$ bits. In the CA literature, the truth table of a local rule is usually represented by its \emph{Wolfram code}~\cite{wolfram83}, which corresponds to the decimal encoding of $\Omega_f$. Figure~\ref{fig:nbca} reports an example of CA of length $8$ and diameter $d$. The local rule, defined as $f(x_i, x_{i+1}, x_{i+2}) = x_i \oplus x_{i+2}$, has Wolfram code $90$.

\section{Related Works}
\label{sec:rel-works}
As we mentioned in the Introduction of this paper, cellular automata have a long history of being investigated for cryptographic applications. The reason for this interest is grounded in two main observations. First, the \emph{shift-invariance} property that characterizes CA allows for uniform and efficient hardware implementations: as we recalled in Section~\ref{subsec:ca} the global rule of a CA is completely determined by the parallel application of the same local rule at all sites of the cellular array. Second, depending on the underlying local rule, the dynamic evolution of a CA can be quite complex and unpredictable. To a first approximation, it seems interesting to exploit the dynamics of certain CA when designing a symmetric cipher that follows the confusion and diffusion principles.

However, relying only on the dynamical properties of CA is usually not sufficient to realize a sound cryptographic primitive, since reasonable levels of security often require much more stringent criteria than those studied in the field of dynamical systems, such as sensitivity to initial conditions~\cite{kurka03}. This has been the case with the first proposal to employ CA for cryptographic applications, namely Wolfram's pseudorandom generator~\cite{wolfram85}. Pseudorandom numbers and sequences are crucial in many cryptographic applications, particularly in the case of \emph{Vernam-like} stream ciphers~\cite{stinson18}. In this type of cipher, a short secret key is stretched through a pseudorandom generator, which produces a keystream of the same length as the plaintext. Then, the encryption simply amounts to the bitwise XOR of the plaintext and the keystream. Decryption works symmetrically by computing the XOR between the ciphertext and the keystream, which the receiver can re-create by running again the pseudorandom generator using the same secret key as a seed. Clearly, the security of this system entirely relies on the properties of the pseudorandom generator, with the assumption that it cannot be efficiently predicted by an attacker. Wolfram proposed to use a periodic-boundary CA equipped with the local rule 30 of diameter $d=3$ to implement such pseudorandom generator. Specifically, the idea was to set the secret key as the initial configuration of the CA, and then to evolve it for several iterations. The \emph{trace} of the central cell (that is, the sequence of states assumed by the cells through different time steps) was then used as a keystream to be XORed with the plaintext.

Wolfram argued the security of this generator on the basis of several system-theoretic and statistical criteria. Indeed, rule 30 belongs to the so-called ``class 3'' of local rules in Wolfram's taxonomy, which induce CA with a chaotic dynamics. Moreover, CA equipped with rule 30 are also known to satisfy Devaney's definition of topological chaos~\cite{leporati14}. Still, Wolfram's pseudorandom generator was later found to be vulnerable against two serious attacks. The first one was demonstrated by Meier and Staffelbach~\cite{meier91}, who showed that one can mount a \emph{correlation attack} that is able to efficiently recover the CA initial configuration. Koc and Apohan~\cite{koc97} further showed that rule 30 can be easily approximated by affine functions, and this allows to efficiently invert the CA iteration (and thus again, to easily recover the initial configuration). Martin~\cite{martin08} later showed that the vulnerabilities of Wolfram's generator can be explained in terms of the cryptographic properties of rule 30, when interpreted as a Boolean function of $3$ variables. Specifically, Meier and Staffelbach's attack succeeds because rule 30 is not \emph{first-order correlation immune}, while Koc and Apohan's attack is related to the low nonlinearity of rule 30.

For the reasons above, more recent works focused on improving Wolfram's generator by searching for rules with a larger diameter and a better trade-off of cryptographic properties. Formenti et al.~\cite{formenti14} used the DIEHARD statistical test suite~\cite{marsaglia96} to investigate the quality of pseudorandom sequences produced by CA with local rules of diameter $5$ of the best nonlinearity and correlation immunity. Leporati et al.~\cite{leporati14} focused on the class of \emph{bipermutive} rules (which are known to induce chaotic CA) and performed a combinatorial search of highly nonlinear correlation-immune rules of diameter $5$ and $7$. The three best rules of diameter $5$ that passed all tests of the NIST suite~\cite{bassham10} in that work were later selected to design the nonlinear components of the {\sc CARPenter}~\cite{lakra18} and {\sc Pentavium}~\cite{john20} CA-based stream ciphers.

Other works considered different directions to design pseudorandom generators based on CA. Sipper and Tommasini~\cite{sipper96} were the first to propose the use of \emph{non-uniform} CA to generate pseudorandom sequences, i.e. CA where each cell may use a different local rule to update its state. In particular, they proposed a \emph{cellular programming} technique where a co-evolutionary algorithm is used to evolve the set of rules to be used in the CA. Tomassini et al.~\cite{tomassini01} and Seredynsky et al.~\cite{seredynski04} subsequently applied the cellular programming approach also to the case of two-dimensional CA for designing strong pseudorandom generators.

A further research direction examined CA-based pseudorandom number generators where the cells have non-binary states. The main motivation underlying this approach is of combinatorial nature: for a given diameter, there are many more local rules on a non-binary alphabet than in the classic Boolean case, which gives more possibilities to find interesting local rules for pseudorandom generation. Clearly, this also brings the problem of actually implementing multi-states CA in hardware, an issue which can be easily addressed in the binary case. Bhattacharjee et al.~\cite{bhattacharjee17} considered CA rules on a ternary alphabet with diameter $d=3$, finding one rule that was able to pass several statistical and empirical tests for randomness. Later, Bhattacharjee et al.~\cite{bhattacharjee19} extended the scope of this investigation to CA with a decimal alphabet, finding a few rules whose randomness can compete with some of the best pseudorandom generators proposed in the literature, both based on CA and not.

Finally, a few recent works contemplated additional CA models for pseudorandom generation. For instance, Manzoni et al.~\cite{manzoni18} performed a statistical investigation through the NIST test suite of a Wolfram-like generator based on \emph{asynchronous CA}, where the cells do not update all at the same time. Interestingly, the authors of that work noticed that the quality of the pseudorandom sequences generated by some local rules actually improves by adding a certain amount of asynchrony. Further, the first author of this manuscript investigated a new model of pseudorandom number generators grounded on the use of \emph{orthogonal CA} (OCA)~\cite{mariot21}. OCA have been recently introduced in the literature as a method to generate orthogonal Latin squares, with the original purpose of designing secret sharing schemes~\cite{mariot18a,mariot20}. The rationale for considering OCA also for pseudorandom number generation is that they guarantee a minimum amount of diffusion, since orthogonal Latin squares are related to MDS codes. Moreover, since orthogonal Latin squares represent a permutation over the Cartesian product of their support set, the dynamics of the resulting generator is also reversible. In particular, \cite{mariot21} describes a combinatorial search algorithm to enumerate all linear OCA pairs that can generate sequences with maximum period, a property which is usually sought in good pseudorandom generators.

\section{The CA-XOR Construction} 
\label{sec:constr}
Almost all works discussed in the previous section have one aspect in common: the preminence of the statistical approach to assess the quality of pseudorandom  generators designed by CA. However, it is widely known that the suitability of a pseudorandom generator for cryptographic applications cannot be established only by means of statistical tests~\cite{carlet21,stinson18}. In particular, statistical tests are useful to discard bad generators, but are in general not sufficient to conclude that certain specific attacks will not break them. Few papers such as~\cite{formenti14,leporati14,mariot21} partially address this issue by considering also the cryptographic properties of the CA local rules, in combination with the statistical quality of the pseudorandom sequences that they produce. Nevertheless, such works are a minority in this research field. Moreover, cryptographic criteria of Boolean functions like balancedness, algebraic degree and nonlinearity were introduced for pseudorandom generators models that are different from those based on CA. Except for correlation-immunity and nonlinearity, which have been shown to be linked respectively to the Meier-Staffelbach and the Koc-Apohan attacks~\cite{martin08,leporati14}, there is no systematic way to translate the relevance of a cryptographic property for Boolean functions in a CA-based pseudorandom generator model.

In this paper we follow the opposite direction: instead of searching for Boolean functions that are embedded in a CA-based pseudorandom generator, here we use CA to synthesize Boolean functions with good cryptographic properties. The rationale is that such Boolean functions defined by CA can then be used in more established designs of pseudorandom generators (such as the combiner or the filter model), and whose security is better understood. In particular, we focus on defining a \emph{secondary construction} of Boolean functions, where a CA is used to expand a known function $f$ of $d$ variables used as a CA local rule, and to generate a new function $f^*$ of $n\ge d$ variables.

\begin{figure}[t]
    \centering
    \includegraphics[scale=0.7]{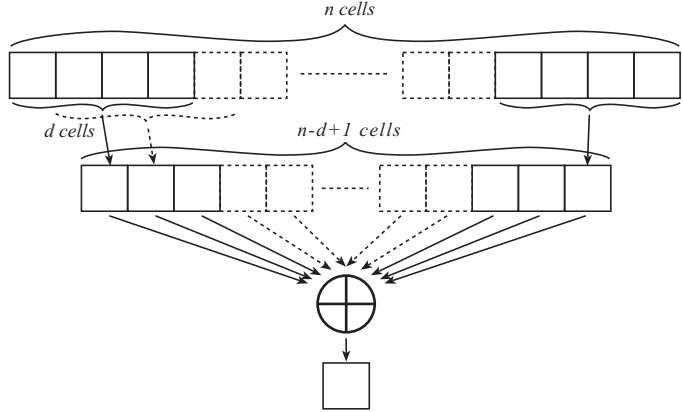}
    \caption{Representation of our CA-based construction for Boolean functions.}
    \label{fig:constr}
\end{figure}

We now give the formal definition of our \emph{CA-XOR construction} for Boolean functions, using the no-boundary CA model introduced in Section~\ref{subsec:ca}.
\begin{definition}
\label{def:constr}
Let $F: \F_2^n \to \F_2^{n-d+1}$ be a CA of length $n\ge d$ equipped with the local rule $f:\F_2^d \to \F_2$ of diameter $d \in \N$. Then, the Boolean function generated by $f$ through the CA $F$ is the $n$-variable function $f^*: \F_2^n \to \F_2$ defined for all $x \in \F_2^n$ as:
\begin{equation}
  \label{eq:constr}
  f^*(x) = \bigoplus_{i=1}^{n-d+1} f(x_i,\cdots,x_{i+d-1}) = f(x_1,\cdots,x_d) \oplus \cdots \oplus f(x_{n-d+1},\cdots,x_n) \enspace .
\end{equation}
\end{definition}
In other words, the CA-XOR construction works by first applying the CA vectorial function $F$ induced by the local rule $f$ to the input vector $x \in \F_2^n$ of $n$ variables; then, the value of the resulting function $f^*$ is obtained by computing the XOR of all the $n-d+1$ output cells of the CA. To illustrate this idea, Figure~\ref{fig:constr} gives a schematic depiction of how the CA-XOR construction.

Secondary constructions are mainly employed to generate new Boolean functions from old ones with analogous cryptographic properties. For example, Rothaus's construction~\cite{rothaus76} starts from three bent functions of $n$ variables, whose sum is also bent, and produces a new bent function of $n+2$ variables. We thus need to analyze which properties are preserved by our construction. The next lemma shows that the algebraic degree is one such property:
\begin{lemma}
\label{lm:alg-deg}
Let $f: \F_2^d \to \F_2$ be a Boolean function of $d$ variables. For any $n\ge d$, the function $f^*$ defined by the CA construction of Equation~\eqref{eq:constr} has the same algebraic degree of $f$.
\begin{proof}
The result is clearly true when $n=d$, since in that case $f \equiv f^*$. We thus only consider the case where $n>m$.

Let $t$ be the algebraic degree of $f$. Each summand in Equation~\eqref{eq:constr} has degree $t$, since it always corresponds to the local rule $f$ applied on a different neighborhood. We thus have to show that not all terms of degree $t$ cancel each other out. Consider the first summand $f(x_1,\cdots, x_d)$, and let $S_t = \{I \subseteq 2^{[d]}: \vert I \vert = t, a_I \neq 0\}$ be the set of monomials of degree $t$ in the ANF of $f$. Further, denote by $I_{min} \in S_t$ the minimum element of $S_t$ with respect to the lexicographic order, that is, if $I_{min} = \{i_1,\cdots, i_t\}$ and $J = \{j_1,\cdots,j_t\}$ is any other set of $S_t$, it holds $i_k < j_k$ for some $k \in [t]$ and $i_h = j_h$ for all $h \in [k-1]$. This monomial cannot be cancelled by any other monomial in the ANF of the subsequent summands, since by Equation~\eqref{eq:constr} their neighborhoods are shifted by at least one coordinate with respect to that of the first summand. Indeed, if we take the $l$-th summand $f(x_{l},\cdots,x_{l+d-1})$ for $l \in \{2,\cdots,n-d+1\}$, and we denote by $I_{min}^l$ its minimum monomial of degree $t$ in lexicographic order, we have that $I_{min}^l = (i_1+l,\cdots,i_t+l)$, which is distinct from $(i_1,\cdots,i_t) = I_{min}$. Hence, the variables in the monomial $I_{min}^l$ cannot overlap completely those of $I_{min}$, which means that the two terms do not cancel each other out. Similarly, the monomial $I_{min}$ cannot be canceled by any non-minimal monomial of degree $d$ in the $l$-th summand. Hence, the monomial corresponding to $I_{min}$ appears in the ANF of~\eqref{eq:constr}, which proves that the algebraic degree of $f^*$ is also $t$.
\end{proof}
\end{lemma}

Remark that the result above does not hold if one considers our construction with  \emph{periodic boundary} cellular automata, where no cells are lost upon application of the local rule. Indeed, the proof of Lemma~\ref{lm:alg-deg} relies on the presence of non-overlapping neighborhoods, that do not cancel each other out in the ANF of the function $f^*$. On the contrary, with periodic boundary conditions one can find several examples where this property is not verified, and where the algebraic degree of the function generated by the CA-XOR construction is strictly less than the degree of the original local rule. This is an additional reason why we sticked to the NBCA model in this paper, and we did not consider evolving the CA for multiple time steps. This not a huge limitation however, since our CA-XOR construction is actually recursive and gives rise to an \emph{infinite family} of Boolean functions: $f^*$ can be defined over any number of variables $n\ge d$ by simply adding $n$ cells to the CA.

\section{Exhaustive Search} 
\label{sec:exhaust}
We now investigate the effectiveness of the CA-XOR construction in building Boolean functions with interesting properties from a cryptographic standpoint. In particular, we are interested in finding, and possibly characterising, a family of Boolean functions from which it is possible to build, by means of our construction, other functions with similar cryptographic properties but with an arbitrary number of variables.

As remarked in Section~\ref{subsec:bf}, there are many properties that need to be taken into account when choosing Boolean functions to be used in the combiner or the filter model. Referring to nonlinearity, bent functions are of particular interest, since they reach the covering radius bound. Nonetheless, given that bent functions in $n$ variables exist only if $n$ is even, and that they are not balanced, in this study we also consider \emph{semi-bent} functions, since they exist also for odd number of variables, they reach the quadratic bound of nonlinearity, and they can also be balanced. More precisely, in the rest of this paper we focus on the search of (semi-)bent Boolean functions that, when plugged into our CA-XOR construction, result in a bent function $f^*$ when the number $n$ of CA cells is even, and in a semi-bent function when $n$ is odd. In this section we present a combinatorial algorithm to exhaustively search for such functions up to diameter $7$, while in the next one we will use evolutionary algorithms to search them for larger diameters.

Lemma~\ref{lm:alg-deg} states that the functions yielded by the CA-XOR construction in Equation~\ref{eq:constr} have the same algebraic degree of the initial function. This remark is especially interesting when considering the case of quadratic functions. Indeed, quadratic functions are a proper subset of plateaued functions~\cite{carlet21}, which in turn include both bend and semi-bent functions, as mentioned in Section~\ref{subsec:bf}.

Moreover, Theorem~\ref{thm:equiv-quad} shows that there exists only a single (semi-)bent quadratic function up to affine equivalence. Therefore, it is of interest to classify which quadratic functions can be successfully extended by the CA-XOR construction under coarser equivalence relations.

On account of Lemma~\ref{lm:alg-deg}, in~\cite{mariot20a} we devised an algorithm that exhaustively visits all algebraic normal forms of degree $d$, retrieve their truth tables via the M\"{obius} transform, and then apply our construction to see which of these functions always produce semi-bent functions up to $n=20$ cells. Considering the case of quadratic functions, the resulting search space explored by the {\sc Search-ANF} algorithm had the following size with respect to the diameter of the starting local rule:

\begin{equation}
\label{eq:search-space}
S_{d,2} = \left(2^{\binom{d}{2}}-1\right)\cdot 2^{\binom{d}{1}} = \left(2^{\frac{d(d-1)}{2}}-1\right)\cdot 2^{d} \enspace .
\end{equation}


In our previous work~\cite{mariot20a}, we applied this algorithm up to diameter $d=6$, where there are approximately 2 millions quadratic functions to explore. Here, we refine our search approach by leveraging on Theorem~\ref{thm:equiv-quad}. As one can see from Eqs.~\eqref{eq:bent-quad} and~\eqref{eq:sbent-quad}, the unique (semi-)bent quadratic function that exists up to affine equivalence is \emph{homogeneous}, meaning that its ANF does not have any linear term. Thus, it makes sense to limit our exhaustive search only to homogeneous quadratic ANFs, since this reduces the search space size. Clearly, linear terms do not influence the nonlinearity of the functions resulting from our CA-XOR construction, which is the main property one aims to preserve when considering (semi-)bent functions. In particular, the nonlinearity of a quadratic function is determined only by the monomials of degree 2 in its ANF.

Our search algorithm is based on the ANF representation. In general, given a target algebraic degree $t$, the $2^d$-bit vector of the ANF coefficients can be easily constrained to yield only Boolean functions of degree $t$: it suffices to set \emph{at least one} of the coefficients $a_I$ such that $\vert I \vert =t$ to $1$, while all coefficients $a_J$ with $\vert J \vert >t$ must be set to 0. Moreover, since we consider only \emph{homogeneous functions}, also the coefficients related to monomials of degree lower degree than $t$ can be set to $0$. Then, by using the \emph{M\"{o}bius Transform} recalled in Equation~\eqref{eq:mobius}, one can recover the truth table starting from its ANF coefficients, and check if the corresponding function is bent or semi-bent by computing its Walsh spectrum. In this case, we can finally test if our construction generates quadratic bent or semi-bent functions (depending on whether the number of variables $n$ is even or odd) up to a specified number of variables.

Algorithm~\ref{alg:search} reports the pseudocode of the simplified search procedure, adapted to the enumeration of quadratic homogeneous ANFs.
\begin{algorithm}[t]
{\sc Search-Quad-ANF}$(d,n)$

\setlength{\parindent}{1em}
{\bfseries Initialization:}
\begin{enumerate}
\setlength{\itemindent}{2em}
\item Build the family $\mathcal{I}_2=\{I \subseteq [d]: \#I = 2\}$ of monomials of degree $2$
\item Set all $2^m$ ANF coefficients of $f$ to 0
\end{enumerate}

\setlength{\parindent}{1em}
{\bfseries Loop:} For all subsets $\mathcal{T} \subseteq \mathcal{I}_2$ (except the empty set), do:
  \begin{enumerate}
  \setlength{\itemindent}{2em}
    \item Reset all $2$-degree terms in the ANF to 0
    \item For all $T \in \mathcal{T}$, set the ANF coefficient $a_T$ to $1$
    \item Set $f$ to the M\"obius Transform (Equation~\eqref{eq:mobius}) of the ANF 
    \item Compute the Walsh transform (Equation~\eqref{eq:walsh}) of $f$
    \item If $f$ is (semi-)bent, then apply the CA-XOR construction up to $n$ cells
    \item If $f$ satisfies the target, print it
  \end{enumerate}
\caption{Pseudocode of the simplified exhaustive search algorithm.}
\label{alg:search}
\end{algorithm}
The {\sc Search-Quad-ANF} algorithm takes as input only the diameter $d$ of the starting local rule and the number of cells $n$ up to which to test the CA-XOR construction. The initialization step consists in constructing the family of all subsets of $2$ variables out of $d$, and then set all coefficients in the ANF vector to zero. Next, the main loop performs the enumeration of all $2^{\binom{d}{2}}-1$ non-empty subsets of of two variables. For each of them, the algorithm set the corresponding ANF coefficients to $1$, then recover the truth table of the function via the M\"obius Transform and checks whether the function is bent or semi-bent, depending on the parity of $d$. If $d$ is even, then the target is to check that $f$ is a bent function, and that by applying the CA-XOR construction for all $d < i \le n$ the resulting function is bent when $i$ is even, and semi-bent when $i$ is odd. If this is the case, then the starting function $f$ is printed.

Remark that {\sc Search-Quad-ANF} is simpler than the {\sc Search-ANF} algorithm described in~\cite{mariot20a}, since only quadratic monomials are considered. More precisely, {\sc Search-ANF} takes as input the number of variables $d$ and the target degree $t$, and then performs two nested loops over all possible sets of monomials having degree \emph{at most} $t-1$ (inner loop), and all possible non-empty sets of monomials of degree \emph{exactly} $t$ (outer loop). Dealing with homogeneous functions, only the outer loop is needed for our purposes. Since functions in $d$ variables have at most $\binom{d}{2}$ terms of degree $2$ in their ANFs, such loop is executed $2^{\binom{d}{2}} - 1$ times, and thus it is feasible for functions up to $7$ variables. For comparison, Table~\ref{tab:sizes} displays the size of the spaces $\mathcal{Q}_d$ and $\mathcal{H}_d$, respectively visited by the {\sc Search-ANF} and the {\sc Search-Quad-ANF} algorithms.

\begin{table}[t]
\caption{Search space sizes for the set of all quadratic functions $\mathcal{Q}_d$ and the set of all homogeneous quadratic functions $\mathcal{H}_d$ of $d$ variables.}
    \centering
    \begin{tabular}{crr}
        \toprule
        $d$ & $\mathcal{Q}_d$   &$\mathcal{H}_{d}$ \\
        \midrule
        3 & 56 & 7 \\
        4 & 1008 & 63 \\
        5 & 32736 & 1023 \\
        6 & 2097088 & 32767 \\
        7 & 268435328 & 2097151 \\
        8 & 68719476480 & 268435455 \\
        \bottomrule
    \end{tabular}
    \label{tab:sizes}
\end{table}

\section{Evolutionary Search} 
\label{sec:es}
As it can be seen from Table~\ref{tab:sizes}, the size of the search space for the set of quadratic functions quickly grows for small sizes already, even by considering only the subset of homogeneous ANFs. For this reason, we enlarged our empirical search of quadratic (semi-)bent functions that our CA-XOR construction can extend by using \emph{Evolutionary Algorithms} (EA).

EA and other nature-inspired optimization heuristics such as swarm intelligence algorithms have been extensively used to optimize the cryptographic properties of Boolean functions. Examples include \emph{Genetic Algorithms} (GA)~\cite{millan98,mariot15,manzoni20}, \emph{Genetic Programming} (GP)~\cite{hcastro03,hrbacek14,picek15} and \emph{Particle Swarm Optimization} (PSO)~\cite{saber06,mariot15b}. Most of these works usually focus on the use of EA and swarm intelligence algorithms as alternative constructions of Boolean functions with good cryptographic properties. In particular, such methods do not work like primary or secondary constructions, since they blindly search only on the basis of a \emph{fitness function} to be optimized, which measures a combination of some properties of interest. Consequently, it is difficult to compare the functions obtained with these methods with those produced by a primary or secondary construction. Moreover, directly employing EA to find quadratic functions of diameter $d>7$ that our CA-XOR construction can extend presents another pratical problem: namely, one should adopt a \emph{two-stage} optimization strategy where first EA are used to evolve (semi-)bent functions. Then, in the second phase the EA should tweak the evolved functions (without decreasing their nonlinearity) so that when plugged into a no-boundary CA, the CA-XOR construction can successfully extend them up to a large number of variables.

\subsection{Candidate Solutions Encoding}
\label{subsec:enc}
Consequently, it would be useful to find a representation of candidate solutions such that the EA is constrained to explore only (semi-)bent quadratic functions, in order to focus the optimization effort only on maximizing the number of cells in the CA-XOR construction. To this end, recall from Theorem~\ref{thm:equiv-quad} that there exists only one bent (respectively, semi-bent) quadratic function of $d$ variables for all $d \in \N$, which we defined as $b(x)$ in Eq.~\eqref{eq:bent-quad} (respectively, $s(x)$ in Eq.~\eqref{eq:sbent-quad}). Since each bent quadratic function can be obtained by applying a suitable affine map to b(x) (or s(x) if we consider the semi-bent case), our basic idea to apply EA to this problem is to \emph{evolve a population of affine transformations}, rather than evolving a population of specific Boolean functions as done in the majority of the works in this area.

Given the number of variables $d$ of the starting Boolean function, the genotype of an individual in the EA population is thus a pair $c = (M, v)$, where $M$ is a $d\times d$ invertible Boolean matrix, while $v \in \F_2^d$ is a vector of $d$ bits. Then, the phenotype corresponding to $c$ is the function $f:\F_2^d \to \F_2$ defined as follows:
\begin{equation}
\label{eq:phen}
f(x) =
\begin{cases}
b(Mx^\top \oplus v), & \textrm{ if } $d$ \textrm{ is even} \enspace , \\
s(Mx^\top \oplus v), & \textrm{ if } $d$ \textrm{ is odd} \enspace , \\
\end{cases}
\end{equation}
where the parity of $d$ is used to understand whether we need to start from a bent function or semi-bent function when $d$ is even or odd, respectively.

\subsection{Mutation Operator}
\label{subsec:mut}
Beside the encoding of the candidate solutions, an EA needs also to specify a set of \emph{variation operators} that are used to create new solutions from the current populations. In this case, we chose to employ \emph{Evolutionary Strategies} (ES), since they only need to define a \emph{mutation operator}. As a matter of fact, other EA such as genetic algorithms also define \emph{crossover operators}, which take in input two individuals in the population and output one or more \emph{offspring} solutions. However, defining a crossover operator that preserves our encoding for the candidate solutions is not straightforward. Indeed, while the vectors $v$ defining the translation of an affine map is easy to handle with classic operators (such as one-point crossover), the same cannot be said for the linear transformation part. In fact, to define a crossover operator between two parents $p_1 = (M,v)$ and $p_2 = (M',v')$, one would need to devise a method to combine the matrices $M, M'$ such that the result is still an invertible matrix. For this reason, we decided to focus only on mutation, which applies a random small tweak to the genotype of a single individual, rather than combining those of two parents.

Mutation of the translation vector $v$ is easy to perform: one can apply the classic mutation operators developed in the EA literature for bitstring-based encodings. In particular, we adopted a simple \emph{bit-flip} operator which, for all positions $i \in [d]$, complements the bit $v_i$ with mutation probability $p_{p_{\mu}} \in (0,1)$. The mutation of the linear map part of an individual, namely the invertible matrix $M$, starts from the following observation. Let $M_i$ be the $d\times d-1$ matrix obtained by removing column $i \in [d]$ from an invertible $d\times d$ Boolean matrix $M$. Since $M$ is invertible, $M_i$ must have rank $d-1$. Hence, the linear transformation $L: x \mapsto Mx^\top$ generates the whole vector space $\F_2^d$, while the map $L': x \mapsto M_i x^\top$ spans a subspace $S_i \subseteq \F_2^d$ of dimension $d-1$. Let $C_i = \F_2^d \setminus S_i$ be the complementary set of the subspace $C_i$. By definition of subspace spanned by the matrix $M_i$, each vector in $C_i$ will be linearly independent with each vector in $S_i$. Therefore, if we take any vector $c \in C_i$ and add it to $M_i$, we obtain an invertible $d\times d$ matrix.

In other words, our mutation operator works by replacing with probability $p_{\mu}$ each column of a matrix with a random vector chosen from the complementary set of the span generated by the remaining columns. In this way, the rank (and thus the invertibility) of the matrix is preserved. Algorithm~\ref{alg:mutation} summarizes our operator as high-level pseudocode.

\begin{algorithm}[t]
{\sc Mutation-Affine}$(M,v,d,p_{\mu})$

\setlength{\parindent}{1em}
{\bfseries Vector part:}

\setlength{\parindent}{2em}
{\bfseries Loop:} for all $i \in [d]$ do:
\begin{enumerate}
\setlength{\itemindent}{3em}
\item Draw a random number $r \in (0,1)$
\item If $r < p_{\mu}$, then flip the bit $v_i$
\end{enumerate}

\setlength{\parindent}{1em}
{\bfseries Matrix part:}

\setlength{\parindent}{2em}
{\bfseries Loop:} for all $i \in [d]$ do:
\begin{enumerate}
\setlength{\itemindent}{3em}
\item Draw a random number $r \in (0,1)$
\item If $r < p_{\mu}$, then:
\begin{enumerate}
\setlength{\itemindent}{4em}
\item Remove the $i$-th column of $M$
\item Generate the span $S_i$ of the remaining columns
\item Randomly pick $c$ from $C_i = \F_2^d \setminus S_i$ and add it to $M$
\end{enumerate}
\end{enumerate}
\caption{Pseudocode of the mutation operator used in our ES algorithm.}
\label{alg:mutation}
\end{algorithm}

\subsection{Overall ES Algorithm Structure}
\label{subsec:es-alg}
There exist two main types of Evolutionary Strategies~\cite{luke11}: in $(\mu,\lambda)$ ES, a population of $\lambda$ initial individuals is first generated at random, and then their fitness is assessed. The $\mu$ best individuals are selected for reproduction. In particular, each of the $\mu$ selected individuals will generate $\lambda/\mu$ new individuals through mutation, thereby producing a new population of size $\lambda$. The ES process is then iterated by assessing the fitness of the new $\lambda$ individuals, selecting the $\mu$ best ones, and applying the mutation operators on them, and so on. The $(\mu+\lambda)$ ES differs only in the fact that the $\mu$ best individuals from the old population are added to the new population of $\lambda$ individuals generated through mutation. Therefore, in this case the parents directly compete also with their children.

For our experiments, we employed both $(\mu,\lambda)$ and $(\mu+\lambda)$ ES. Since the encoding of the candidate solutions decribed in Section~\eqref{subsec:enc} guarantees that the phenotypes of the individuals are all (semi-)bent quadratic functions, the fitness function used in our ES optimizes another criterion. In particular, each function $f$ in the population is used as a local rule of diameter $d$ of a no-boundary CA. Given a target number of cells $N \in \N$, the fitness function applies the CA-XOR construction for all CA lengths $d < n \le N'$, where $N'\le N$ is the maximum number of cells up to which the construction always yields bent functions when $n$ is even, and semi-bent functions when $n$ is odd. The fitness value of $f$ is then $N'-d$, i.e. the number of cells over which the CA-XOR construction successfully works when using $f$ as a local rule.

\section{Experiments}
\label{sec:class}

We applied the exhaustive search algorithm described in Section~\ref{sec:exhaust} on quadratic functions of $3 \leq d \leq 7$ variables, while we applied the ES approach for functions of $8 \le d \le 12$ variables. In the case of exhaustive search, we tested our CA-based construction up to $N=20$ cells. On the other hand, with ES the fitness function applied the CA-XOR construction to a maximum of $N=16$ cells. This is due to the fact that testing the construction up to $20$ cells for a single starting function of more than $7$ variables takes approximately 20 seconds on the server that we used to perform our experiments, which is equipped with a Core i7 processor running at 2.8 GHz. Since the ES algorithm needs to test several thousands of these functions during the evolution process, the amount of time required to finish a single optimization run would have been to large. For the other parameters related to the ES, we set $\lambda = d$ and $\mu = \lceil \lambda / 3 \rceil$. Therefore, the diameter of the local rules determines the number of individuals generated at each iteration. The mutation probability was set to $p_{\mu} = 1/\lambda$, a common choice in the field of GA and ES~\cite{luke11}. Further, for each considered diameter $8 \le d \le 12$ we performed $100$ independent runs of $(\mu,\lambda)$ and $(\mu+\lambda)$ ES, giving a budget of $1000$ generations for each run.

To investigate the results of our exhaustive search and ES experiments, we classified the obtained functions up to permutation equivalence. As a matter of fact, we already know by Theorem~\ref{thm:equiv-quad} that there exists only a single (semi-)bent quadratic function up to affine equivalence, hence a classification based on this equivalence relation would not give us any insight into the functions found in our search experiments. Consequently, we needed to a coarser relation to discriminate different groups of functions that our CA-XOR construction can successfully extend.

\subsection{Exhaustive Search Results}
\label{subsec:ex-res}
The results for the exhaustive enumeration algorithm {\sc Search-Quad-ANF} are summarized in Table~\ref{tab:exhaustive}. In particular, for each considered diameter $d$, we report in column $\mathcal{H}_d$ the total number of homogeneous quadratic functions in $d$ variables, that is $2^{\binom{d}{2}} - 1$. Columns $\mathcal{C}_{16}$ and $\mathcal{C}_{20}$ show the total number of functions over which our construction works up to $16$ and $20$ cells respectively, namely:

\begin{itemize}
    \item bent functions in an even number of variables that always produces bent functions when an even number of variables is added to the CA and semi-bent functions when an odd number of variables is added to the CA;
    
    \item semi-bent functions in an odd number of variables that always produces semi-bent functions when an even number of variables is added to the CA and bent functions when an odd number of variables is added to the CA.
\end{itemize}

In other words, we are looking for Boolean functions that always yield bent functions when $n$ is even and semi-bent functions when $n$ is odd. The last column reports the number of distinct classes represented among the functions for which the construction works up to $n = 20$ cells, up to permutation equivalence.

\begin{table}[tb]
\caption{Results obtained with the exhaustive search experiments. We denote with $\mathcal{H}_d$ the total number of homogeneous quadratic functions in $d$ variables, and with $\mathcal{C}_{n}$ the number of functions for which the construction applied up to $n$ cells always produces bent or semi-bent functions (for $n$ even or odd, respectively).}
    \centering
    \begin{tabular}{crrrr}
        \toprule
        $d$ &$\mathcal{H}_d$   &$\mathcal{C}_{16}$ &$\mathcal{C}_{20}$     &classes\\
        \midrule
        3   &7                  &0                  &0                      &0\\    
        4   &63                 &3                  &3                      &1\\
        5   &1\,023             &22                 &22                     &7\\
        6   &32\,767            &318                &318                    &44\\
        7   &2\,097\,151        &15\,656            &10\,974                &797\\
        \bottomrule
    \end{tabular}
    \label{tab:exhaustive}
\end{table}

One can see from the table that the number of distinct permutation classes grows at increasing diameter. Namely, one starts from the smallest instance $d=3$ where no quadratic semi-bent functions can be extended by the CA-XOR construction up to $n=16$ cells (let alone $n=20$ cells), while for diameter $d=7$ one already obtains $797$ distinct classes. It is also of interest to notice that for $d=7$ the number of functions for which the construction works up to 20 cells is lower than for 16 cells, while for lower diameters it is always equal. This might indicate that, as the diameter grows, there are more (semi-)bent quadratic functions for which the CA-XOR construction works up to a certain point, but then fails for all larger CA lengths.

\subsection{Evolutionary Strategies Results}
\label{subsec:es-res}
The results obtained by our $(\mu,\lambda)$ and $(\mu+\lambda)$ ES are somehow surprising. The first remarkable finding is that, for all considered diameters $8 \le d \le 12$, both ES algorithms \emph{always managed to converge to an optimal solution in all $100$ experimental runs}. In other words, the two algorithms always find  (semi-bent) quadratic functions that the CA-XOR construction can successfully extend up to $N=16$ cells. This is especially interesting considering the limited computational budget of $1000$ generations given to both algorithms.

Consequently, for each diameter we obtained a set of $200$ functions that can be extended up to $16$ cells. We filtered these solutions by checking which among them could also be extended to $20$ cells under the CA-XOR construction. For each diameter, this step always resulted in an average of 30 functions, which is in line with the observation done for diameter $d=7$ in the previous section: for larger diameters, there exist many (semi-)bent quadratic functions that can be extended up to a certain point, after which our construction always fails.

Finally, we classified the filtered functions of diameter $d=8$ and $d=9$ up to permutation equivalence. As a matter of fact, to check if two functions of $d$ variables are permutation equivalent one needs to find a suitable permutation among all the $d!$ possible ones. Considering that we had an average of $30$ functions to classify, this task was feasible only up to $d=9$ variables. After the classification step, we observed the the remarkable fact that \emph{each function belongs to a different equivalence class}. This remark seems to confirm what we already observed in the previous section with the exhaustive search results, namely that the number of permutation classes grows quickly with respect to the CA diameter. Moreover, our ES algorithms seem to be able to uniformly sample among the distribution of such classes, instead of getting stuck in the same local optima. This is a reasonable question, since having an evolutionary algorithm that always converges over all experimental runs could indicate that it always finds the same solution. However, this doubt is cleared by the fact that we obtained as many permutation equivalence classes as many functions in our classification.

\section{Conclusions}
\label{sec:outro}
In this paper, we continued the investigation of a secondary construction of Boolean fuctions based on cellular automata that was initially introduced in~\cite{mariot20a}. The CA-XOR construction uses a Boolean function $f: \F_2^d \to \F_2$ as a local rule of diameter $d$ in a no-boundary CA of $n \ge d$ cells, and defines a new function $f^*:\F_2^n \to \F_2$ of $n$ variables by computing the XOR of the CA ouput cells once the local rule $f$ has been evaluated over the input vector. A preliminary theoretical investigation shows that such construction preserves the algebraic degree of the starting function used as a local rule, which motivates the focus on quadratic functions. Indeed, quadratic functions have a well-known structure, and it is known that there exist only one (semi-)bent quadratic function up to affine equivalence. Therefore, it is interesting to find which quadratic functions can be successfully extended by the CA-XOR construction, using coarser equivalence relations such as permutation equivalence. To this end, we refined the combinatorial algorithm proposed in ~\cite{mariot20a} to enumerate only homogeneous quadratic ANFs, which allowed us to exhaustively search all (semi-)bent quadratic functions that our construction can extend up to diameter $d=7$. Further, we devised a $(\mu,\lambda)$ and $(\mu+\lambda)$ Evolutionary Strategies algorithm to evolve functions that can be extended by the CA-XOR construction for diameters $8 < d \le 12$. The results of our search experiments show that the number of permutation equivalence classes quickly grows with respect to the diameter of the CA local rule. In particular, beside observing a growing number of distinct classes in the exhaustive search experiments, this observation is also corroborated by the fact that the functions found by our ES algorithms all belong to different classes.

The present work is far from exhausting all aspects related to the CA-XOR construction, and there are several avenues for further research on the topic. The first direction worth exploring is a deeper investigation of the equivalence classes found in our heuristic search experiments. In particular, it would be interesting to give a theoretical characterization of the equivalence classes that the CA-XOR can extend \emph{recursively}, i.e. no matter the CA length one always obtain a (semi-)bent function. A possible way to approach this problem is to use the \emph{graph representation} described in~\cite{tokareva15}. In particular, the graph of a quadratic bent function is defined as the graph where the nodes are the input variables, and two nodes are connected by an edge if and only if the multiplication of the corresponding variables occurs as a quadratic monomial in the ANF of the function. A theoretical characterization of the functions which can be extended by the CA-XOR construction might then be inferred by the properties of such graphs.

A second idea to explore concerns the generalization of the CA-XOR construction. In particular, one could consider variations where not all output cells in the CA are XORed to define the constructed function, but only a subset of them. Notice that this equivalent to study the \emph{component functions} of the S-box defined by the no boundary CA. As a matter of fact, the interest for studying the CA-XOR construction originated from the need of characterizing the cryptographic properties of S-boxes defined by CA. The authors of~\cite{mariot19} recently showed an upper bound on the nonlinearity of such S-boxes in terms of the nonlinearity of the underlying local rule. The CA-XOR construction studied in this paper can be seen as a single component of a CA-based S-box, namely the linear combination that sums all output coordinates. Consequently, studying under which conditions a generalized CA-XOR construction yields Boolean functions with a specific value of nonlinearity could help in deriving new bounds for CA-based S-boxes. 

Finally, a very interesting research direction would be to investigate the CA-XOR construction with functions of higher algebraic degree, such as cubic bent functions. In this regard, we performed a preliminary test on the cubic semi-bent functions of $6$ variables found in~\cite{mariot15} through a genetic algorithm. However, none of them could be successfully extended by our CA-XOR construction. This remark calls for a broader and more systematic evaluation, by repeating the exhaustive search experiments performed in this paper also for cubic functions, up to a number of variables where this is still feasible. 

\bibliographystyle{abbrv}
\bibliography{bibliography}

\end{document}